\documentclass[pre,aps,preprint,showpacs,preprintnumbers,amsmath,amssymb,secnumroman,eqsecnum]{revtex4}

\usepackage{graphicx}
\usepackage{dcolumn}
\usepackage{bm}

\newcommand{\beq}{\begin{equation}}
\newcommand{\eeq}{\end{equation}}
\newcommand{\beqa}{\begin{eqnarray}}
\newcommand{\eeqa}{\end{eqnarray}}
\newcommand{\vc}[1]{\mbox{\boldmath $#1$}}

\newcommand{\vol}[1]{{\bf #1}}

\begin{document}


\title{Generalized Einstein relation for the mutual diffusion coefficient of a binary fluid mixture}

\author{B. U. Felderhof}

 \email{ufelder@physik.rwth-aachen.de}
\affiliation{Institut f\"ur Theorie der Statistischen Physik \\ RWTH Aachen University\\
Templergraben 55\\52056 Aachen\\ Germany\\
}%

\date{\today}

\begin{abstract}
The method employed by Einstein to derive his famous relation between the diffusion coefficient and the friction coefficient of a Brownian particle is used to derive a generalized Einstein relation for the mutual diffusion coefficient of a binary fluid mixture. The expression is compared with the one derived by de Groot and Mazur from irreversible thermodynamics, and later by Batchelor for a Brownian suspension. A different result was derived by several other workers in irreversible thermodynamics. For a nearly incompressible solution the generalized Einstein relation agrees with the expression derived by de Groot and Mazur. The two expressions also agree to first order in solute density. For a Brownian suspension the result derived from the generalized Smoluchowski equation agrees with both expressions.
\end{abstract}

\maketitle
\section{\label{I}Introduction}

In his theory of Brownian motion Einstein \cite{1} derived the famous relation between the diffusion coefficient and the friction coefficient of a Brownian particle. For the latter he used the formula derived by Stokes for friction in an incompressible viscous fluid. The result is known as the Stokes-Einstein relation. In a second article Einstein \cite{2} provided a more elaborate argument based on the osmotic pressure of the solution and a simple picture in which an equilibrium system, made non-uniform by an applied potential, is viewed as a situation with two canceling dissipative currents. Einstein applied the argument to a dilute solution, and in this way rederived his earlier result.

In the following we show that Einstein's argument \cite{2} can be used to find the mutual diffusion coefficient of a binary fluid mixture. Other workers made efforts to generalize Einstein's relation to dense solutions, but did not use his argument. Prigogine \cite{3} attempted a generalization on the basis of the thermodynamics of irreversible processes, but his result is indefinite, since he did not precisely specify the thermodynamic derivative which occurs in his relation.

A precise relation was derived by a similar method by de Groot and Mazur in their monograph on non-equilibrium thermodynamics \cite{4}. They argued that the process of diffusion is isothermal and isobaric. As a consequence they find that the diffusion coefficient is proportional to a thermodynamic derivative taken at constant temperature and pressure. The de Groot-Mazur result differs from the generalized Einstein relation which we find in Sec. II on the basis of Einstein's argument. The two expressions agree for a semi-dilute solution and a nearly incompressible mixture.

Batchelor \cite{5} presented a different derivation for a Brownian suspension of hard spheres. As we discussed earlier \cite{6}, his derivation of the thermodynamic derivative is problematic. His expression for the diffusion coefficient agrees with that of de Groot and Mazur. Schurr \cite{6A} used Einstein's argument to derive an expression for the diffusion coefficient which he took to be identical to Batchelor's, but which actually is another way of writing the generalized Einstein relation, as discussed in Sec. IX. Several other workers \cite{7}-\cite{9} derived a third expression from irreversible thermodynamics.

For semi-dilute solutions the mutual diffusion coefficient $D$ is expanded in powers of the number density of the solute. The virial coefficient $k_D$ of the term linear in density is the same for the de Groot-Mazur expression and the generalized Einstein relation, but differs for the third expression \cite{7}-\cite{9} for $D$. The third expression was ruled out by experimental determination of the coefficient $k_D$ for a suspension of silica particles \cite{9A}.

The statistics of solute particle positions of a Brownian suspension can be studied on the basis of the generalized Smoluchowski equation. The latter employs a reduced description in which the solvent is treated as a continuum. The equation holds on a long time-scale on which particle velocities have thermalized and can be left out of consideration. The solvent is not necessarily incompressible. The many-particle Smoluchowski equation was formulated first by Deutch and Oppenheim \cite{10},\cite{11}.

In earlier work \cite{12} we derived an expression for the virial coefficient $k_D$ from the generalized Smoluchowski equation. The method allows calculation for general direct and hydrodynamic pair interactions. For a semi-dilute suspension of hard spheres our result agreed with that of Batchelor \cite{5}. The derivation was generalized to two species of Brownian particles by Jones \cite{13}.

As we show below, for a Brownian suspension of interacting particles the diffusion coefficient found from the generalized Smoluchowski equation agrees with both the de Groot-Mazur expression and that found from the generalized Einstein relation, also for a compressible solvent. For a binary fluid mixture the generalized Einstein relation yields a diffusion coefficient which in general differs from that of de Groot and Mazur \cite{4}. It would be of interest to compare the two expressions in computer simulation and experiment for binary mixtures of particles of comparable size at intermediate densities.

\section{\label{II}Generalized Einstein relation}

We consider a binary fluid mixture of two species of particles, labeled 0 and 1. We denote species 0 as the solvent and species 1 as the solute. We are concerned with mutual diffusion of the two species. We assume that thermal diffusion is fast on the time scale of particle diffusion, so that the temperature $T$ may be regarded as constant. Our goal will be to derive an expression for the mutual diffusion coefficient $D$ in terms of a thermodynamic derivative and a coefficient of mutual friction, generalizing the relation derived by Einstein \cite{1} for a Brownian particle.

In a vessel of volume $V$ of sufficiently large size, and in the absence of any one-body potentials acting on the particles, the mixture in thermal equilibrium is spatially uniform on the macroscopic length scale
with number densities $ n_0$ and $n_1$. Equilibrium thermodynamics of the system is contained in the Helmholtz free energy function
\begin{equation}
\label{2.1}F(n_0,n_1,T,V)=V\varphi(n_0,n_1,T),
\end{equation}
where $\varphi(n_0,n_1,T)$ is the Helmholtz free energy per unit volume. It is assumed that this is a known function of $n_0,\;n_1$ at the given temperature $T$. The corresponding chemical potentials are
\begin{equation}
\label{2.2}\mu_0(n_0,n_1,T)=\frac{\partial\varphi}{\partial n_0},\qquad\mu_1(n_0,n_1,T)=\frac{\partial\varphi}{\partial n_1}.
\end{equation}
The Gibbs-Duhem relation reads
\begin{equation}
\label{2.3}\varphi+p=n_0\mu_0+n_1\mu_1,
\end{equation}
where $p=p(n_0,n_1,T)$ is the pressure. In differential form at constant $T$
\begin{equation}
\label{2.4}dp=n_0d\mu_0+n_1d\mu_1.
\end{equation}

Following Einstein \cite{2} we perform a Gedankenexperiment in which the particles of species 1 are subjected to a spatially varying one-body potential $\Phi_1(\vc{r})$. As a consequence, in thermal equilibrium the number densities $n_0(\vc{r})$ and $n_1(\vc{r})$ also become non-uniform. If $\Phi_1(\vc{r})$ differs from zero only in a macroscopic subvolume $V_0$ with $V_0<<V$, then the density variations are limited to $V_0$, apart from a boundary layer with thickness of the order of the correlation length $\xi$. At large distance from the subvolume the chemical potentials
are uniform and still equal to the equilibrium values $\mu^0_0$ and $\mu^0_1$. On the macroscopic length scale the potentials satisfy
\begin{equation}
\label{2.5}\mu_0(n_0(\vc{r}),n_1(\vc{r}),T)=\mu^0_0,\qquad\mu_1(n_0(\vc{r}),n_1(\vc{r}),T)+\Phi_1(\vc{r})=\mu^0_1.
\end{equation}
We write the number denities as
\begin{equation}
\label{2.6}n_0(\vc{r})=n^0_0+n^1_0(\vc{r}),\qquad n_1(\vc{r})=n^0_1+n^1_1(\vc{r}).
\end{equation}
To first order in the deviations $n_0^1, n_1^1$ the equilibrium conditions Eq. (2.5) become
\begin{eqnarray}
\label{2.7}\frac{\partial\mu_0}{\partial n_0}n_0^1+\frac{\partial\mu_0}{ \partial n_1}n_1^1=0,\nonumber\\
\frac{\partial\mu_1}{\partial n_0}n_0^1+\frac{\partial\mu_1}{\partial n_1}n_1^1+\Phi_1=0,
\end{eqnarray}
where the derivatives are taken in the state $n_0^0, n_1^0$.
Solving for $n_1^1$ we obtain
\begin{equation}
\label{2.8}\bigg(\frac{\partial\mu_1}{\partial n_1}-\frac{\partial\mu_1}{\partial n_0}\frac{\partial\mu_0/\partial n_1}{\partial\mu_0/\partial n_0}\bigg)n^1_1=-\Phi_1.
\end{equation}
The prefactor may be expressed as
\begin{equation}
\label{2.9}\frac{\partial\mu_1}{\partial n_1}-\frac{\partial\mu_1}{\partial n_0}\frac{\partial\mu_0/\partial n_1}{\partial\mu_0/\partial n_0}=\bigg(\frac{\partial\mu_1}{\partial n_1}\bigg)_{\mu_0}.
\end{equation}
We note the symmetry $\partial\mu_1/\partial n_0=\partial\mu_0/\partial n_1$.

Since we deal with an equilibrium situation the current of species 1 vanishes, so that to first order
\begin{equation}
\label{2.10}\vc{J}^1_1=-D\nabla n^1_1-\frac{n_1^0}{\zeta^*_{10}}\nabla\Phi_1=0,
\end{equation}
where $D$ is the mutual diffusion coefficient and $\zeta^*_{10}$ is the friction coefficient defined from the relative mean particle velocity as
 \begin{equation}
\label{2.11}\zeta^*_{10}(\vc{U}_1-\vc{U}_V)=\vc{E}_1,
\end{equation}
where $\vc{U}_V$ is the mean volume velocity
 \begin{equation}
\label{2.12}\vc{U}_V=n_0\overline{v_0}\vc{U}_0+n_1\overline{v_1}\vc{U}_1,
\end{equation}
with partial volumes $\overline{v_0}$ and $\overline{v_1}$ which may be evaluated from the ratios \cite{6}
\begin{equation}
\label{2.13}\overline{v_j}=\frac{\partial p/\partial n_j}{n_0\partial p/\partial n_0+n_1\partial p/\partial n_1},\qquad (j=0,1).
\end{equation}
The mean volume velocity vanishes in the laboratory frame \cite{14}. In Eq. (2.11) it is assumed that in a non-equilibrium situation with a force $\vc{E}_1$ acting on particles of species 1 the relative velocity is locally proportional to the force. Substituting Eq. (2.8) into Eq. (2.10) we obtain for the mutual diffusion coefficient
\begin{equation}
\label{2.14}D_{GE}=\frac{n_1}{\zeta^*_{10}}\bigg(\frac{\partial\mu_1}{\partial n_1}\bigg)_{T,\mu_0}.
\end{equation}
We call this the generalized Einstein relation.

We remark that the friction coefficient $\zeta^*_{10}$ is to be distinguished from the coefficient of relative friction $\zeta_{10}$ defined by
 \begin{equation}
 \label{2.15}\zeta_{10}(\vc{U}_1-\vc{U}_0)=\vc{E}_1.
\end{equation}
The coefficients are related by
 \begin{equation}
 \label{2.16}\zeta^*_{10}=\frac{\zeta_{10}}{1-\overline{\phi_1}},
\end{equation}
where
 \begin{equation}
 \label{2.17}\overline{\phi_1}=n_1\overline{v_1}=1-n_0\overline{v_0}
\end{equation}
is the volume fraction of the solute.

In Eq. (2.10) it is postulated that the equilibrium situation may be viewed as a state in which a diffusion current is balanced by the current driven by the imposed one-body potential in the laboratory frame where the mean volume velocity vanishes. The two coefficients $D$ and $\zeta^*_{10}$ and the thermodynamic derivative in Eq. (2.14) can be measured independently, so that in principle the generalized Einstein relation can be checked experimentally.

\section{\label{III}Results from irreversible thermodynamics}

In his review of liquid diffusion Onsager \cite{15} does not mention Einstein's work, but he derives a relation for the diffusion coefficient involving a thermodynamic derivative of the chemical potential at constant pressure. Prigogine \cite{3} derives an Einstein relation, but his result is indefinite, since he does not mention which quantities are to be held constant in the thermodynamic derivative. In their monograph on non-equilibrium thermodynamics de Groot and Mazur \cite{4} argue that diffusion processes occur under conditions of constant temperature and pressure. Accordingly from the expression for the entropy production they derive an expression for the diffusion coefficient which can be shown to be similar to Eq. (2.14), but with the thermodynamic derivative taken at constant temperature and pressure. The prefactor is first written formally with an Onsager coefficient (their Eq. (XI.124)). At a later stage they introduce the particle mobility. In their expression Eq. (XI.204) for the diffusion coefficient we can choose the molar description with molar friction coefficient $\zeta^m_{10}$ defined from particle velocity relative to a mean molar velocity
 \begin{equation}
 \label{3.1}\vc{U}^m=\frac{n_0\vc{U}_0+n_1\vc{U}_1}{n_0+n_1}.
\end{equation}
Using $\vc{U}_V=0$ and $n_0\overline{v_0}+n_1\overline{v_1}=1$ we find
 \begin{equation}
 \label{3.2}\zeta^m_{10}=(n_0+n_1)\overline{v_0}\zeta^*_{10}.
\end{equation}
 Their variable $n_1$ is the molar fraction $n_1/(n_0+n_1)$ in our notation. The diffusion coefficient can be expressed as
 \begin{equation}
 \label{3.3}D_{GM}=\frac{n_1}{n_0\overline{v_0}\zeta^*_{10}}\bigg(\frac{\partial\mu_1}{\partial n_1}\bigg)_{T,p}.
\end{equation}
The thermodynamic derivative can be calculated as
 \begin{equation}
 \label{3.4}\bigg(\frac{\partial\mu_1}{\partial n_1}\bigg)_{T,p}=\frac{\partial\mu_1}{\partial n_1}-\frac{\partial\mu_1}{\partial n_0}\frac{\partial p/\partial n_1}{\partial p/\partial n_0}.
\end{equation}
It differs from the one in Eq. (2.9).

Yamakawa's derivation \cite{7} is also based on non-equilibrium thermodynamics. He finds
 \begin{equation}
 \label{3.5}D_Y=\frac{n_1}{\zeta^*_{10}}\bigg(\frac{\partial\mu_1}{\partial n_1}\bigg)_{T,p}.
\end{equation}
His replacement of the thermodynamic derivative by one involving the osmotic pressure implies that he assumes the derivative to be taken at constant temperature and pressure. Fujita \cite{8} also agrees with Eq. (3.5). Berne and Pecora's \cite{9} Eq. (13.5.19) is identical with Eq. (3.5) with use of Eq. (2.16). These various authors relate the diffusion coefficient to friction by a comparison of the particle velocity due to the thermodynamic force $-\nabla\mu_1$ in a density gradient with that due to a force $\vc{E}_1$ acting on the particles of species 1. In his review of the theory of Brownian motion Mazo \cite{16} writes the mutual diffusion coefficient in the form Eq. (3.5), but he does not specify the friction coefficient, so that his expression is consistent with Eq. (3.3) if Eq. (2.16) is used and the friction coefficient interpreted as $\zeta_{10}$.

We note that Einstein's equilibrium situation is not isobaric. In mechanical equilibrium \cite{4} the pressure balance is
 \begin{equation}
 \label{3.6}\nabla p=n_0\nabla\mu_0+n_1\nabla\mu_1=n_1\vc{E}_1.
\end{equation}
To first order this agrees with Eq. (2.5). Here the force on species 1 is balanced by both chemical potential gradients. There are cross-effects due to interactions, and in the isobaric situation one should take account of the gradient $\nabla\mu_0$ in the force balance.

For now we note that the various arguments lead to different results for the mutual diffusion coefficient, as shown in Eqs. (2.14), (3.3), and (3.5). It is easily checked for simple models \cite{6} that the two thermodynamic derivatives differ in general. As mentioned in the Introduction, the expression Eq. (3.5) was ruled out by experiment \cite{9A}. The relation between the expressions Eqs. (3.3) and (3.5) was discussed by Schurr \cite {6A}.

\section{\label{IV}Comparison of expressions}

First we compare the expressions for the thermodynamic derivatives and the diffusion coefficients for the case of a two-component gas by expansion in both number densities. The virial expansion of the pressure of  a two-component gas \cite{17} reads to second order in the two number densities $n_0,n_1$
\begin{equation}
 \label{4.1}p=k_BT[n_0+n_1+B_{20}n_0^2+B_{11}n_0n_1+B_{02}n_1^2]+O(n^3).
\end{equation}
Correspondingly the expansion of the two chemical potentials reads
 \begin{eqnarray}
 \label{4.2}\mu_0&=&g_0+k_BT[\log n_0+2B_{20}n_0+B_{11}n_1]+O(n^2),\nonumber\\
 \mu_1&=&g_1+k_BT[\log n_1+B_{11}n_0+2B_{02}n_1]+O(n^2),
\end{eqnarray}
with constants $g_0,g_1$. Hence we find for the two derivatives by use of Eqs. (2.9) and (3.4)
 \begin{eqnarray}
 \label{4.3}n_1\bigg(\frac{\partial\mu_1}{\partial n_1}\bigg)_{T,\mu_0}&=&k_BT[1+2B_{02}n_1]+O(n^2),\nonumber\\
 n_1\bigg(\frac{\partial\mu_1}{\partial n_1}\bigg)_{T,p}&=&k_BT[1+(2B_{02}-B_{11})n_1]+O(n^2).
\end{eqnarray}
The virial coefficients can be calculated from integrals involving the pair interactions between particles  \cite{17}. In particular, for a mixture of hard spheres of radii $a_0,a_1$ the coefficients are
 \begin{equation}
 \label{4.4}B_{20}=\frac{16\pi}{3}\; a_0^3,\qquad B_{11}=\frac{4\pi}{3}\;(a_0+a_1)^3,\qquad B_{02}=\frac{16\pi}{3}\; a_1^3.
\end{equation}
This shows that the difference of the expressions in Eq. (4.3) is significant. We note that in the dilute limit $n_0\rightarrow 0,n_1\rightarrow 0$ at fixed $x=n_1/n_0$ the ratio of coefficients is
 \begin{equation}
 \label{4.5}\lim_{n\rightarrow 0,x}\frac{D_{GE}}{D_{GM}}=\frac{1}{1+x},
\end{equation}
where we have used that both $\overline{v_0}$ and $\overline{v_1}$ tend to $1/(n_0+n_1)$ in the dilute limit.

More generally, Vrij \cite{18},\cite{19} derived from thermodynamics by use of the Gibbs-Duhem equation,
\begin{equation}
 \label{4.6}n_1\bigg(\frac{\partial\mu_1}{\partial n_1}\bigg)_{T,p}=\frac{1-n_1\overline{v_1}}{n_1(\kappa_{\pi_1}-\kappa)},
\end{equation}
where $\kappa$ is the compressibility of the solution and $\kappa_{\pi_1}$ is the osmotic compressibility,
\begin{equation}
\label{4.7}
\frac{1}{\kappa_{\pi_1}}=n_1\bigg(\frac{\partial p}{\partial n_1}\bigg)_{T,\mu_0}=n_1\bigg(\frac{\partial P_1}{\partial n_1}\bigg)_{T,\mu_0}=n_1^2\bigg(\frac{\partial\mu_1}{\partial n_1}\bigg)_{T,\mu_0},
\end{equation}
with osmotic pressure $P_1$. For a nearly incompressible solution the compressibility $\kappa$ can be neglected, so that in that case we obtain
\begin{equation}
 \label{4.8}\bigg(\frac{\partial\mu_1}{\partial n_1}\bigg)_{T,p}\approx(1-n_1\overline{v_1})\bigg(\frac{\partial\mu_1}{\partial n_1}\bigg)_{T,\mu_0},
\end{equation}
corresponding to
\begin{equation}
 \label{4.9}D_{GM}\approx D_{GE}.
\end{equation}

If the solvent is dense, but the solute is dilute, we can perform a virial expansion in powers of the density of the solute. According to McMillan-Mayer theory \cite{17} the osmotic pressure of the solute is given by a virial expansion of the form
\begin{equation}
 \label{4.10}P_1(n_1,\mu_0,T)=k_BT[n_1+B_2n_1^2]+O(n_1^3),
\end{equation}
with a virial coefficient $B_2$ which can be calculated from the effective pair interaction between two particles of the solute immersed in a solvent of chemical potential $\mu_0$. The corresponding expansion of the chemical potential of the solute is
\begin{equation}
 \label{4.11}\mu_1(n_1,\mu_0,T)=g_1+k_BT[\log n_1+2B_2n_1]+O(n_1^2),
\end{equation}
where in general the first term $g_1$ depends on the solvent chemical potential $\mu_0$.
The expansion yields
\begin{equation}
 \label{4.12}n_1\bigg(\frac{\partial\mu_1}{\partial n_1}\bigg)_{T,\mu_0}=k_BT[1+2B_2n_1]+O(n_1^2),
\end{equation}
whereas
\begin{equation}
 \label{4.13}n_1\bigg(\frac{\partial\mu_1}{\partial n_1}\bigg)_{T,p}\approx k_BT[1+(2B_2-\overline{v_1})n_1]+O(n_1^2),
\end{equation}
by use of Eq. (4.8).

In the work of Berne and Pecora \cite{9} their Eq. (13.5.24a) is incorrect. They use an unconventional notation for the virial coefficient $B_2$, and a term $-\overline{v_1}$ is missing on the left hand side \cite{6A},\cite{19A}. Therefore their Eq. (13.5.27) does not follow from the preceding equations. There is no compensating error, as suggested by Kops-Werkhoven et al. \cite{19}. The expression Eq. (4.13) agrees with Yamakawa's Eq. (30.43). We disagree with the comment of Kops-Werkhoven et al. \cite{19} on his expression for the diffusion coefficient, which is identical with Eq. (3.5) with use of Eq. (4.13). The work of Vink \cite{20} is based on the same force balance as used by Berne and Pecora and leads also to Eq. (3.5).

As we noted earlier \cite{6}, Batchelor \cite{5} presented a flawed derivation of a thermodynamic relation which is valid approximately for a nearly incompressible solution, but does not hold in general. The expression in his Eq. (6.8) for the virial expansion of the chemical potential of the solute is incorrect. Apparently there is an error in the derivation from his Eq. (6.6). The correct expansion reads
\begin{equation}
\label{4.14}\mu_1(n_1,\mu_0,T)=g_1+k_BT\log n_1-k_BT\sum^\infty_{j=1}\beta_jn_1^j,
\end{equation}
with coefficients $\beta_j$, in Hill's notation \cite{17}, which depend on the chemical potential $\mu_0$ of the solvent. In our earlier work \cite{6} we denoted Batchelor's quantity as $\mu_{1B}$ and $\mu_1(n_1,\mu_0,T)$ as $M_1$. The osmotic pressure $P_1$ may be expressed by the virial expansion
 \begin{equation}
\label{4.15}P_1(n_1,\mu_0,T)=n_1k_BT\bigg[1-\sum^\infty_{j=1}\frac{j}{j+1}\beta_jn_1^j\bigg].
\end{equation}
The osmotic pressure $P_1$ and the potential $M_1$ are related by the Gibbs-Duhem equation $dP_1=n_1dM_1$ at constant $T$ and $\mu_0$, so that
 \begin{equation}
\label{4.16}n_1\bigg(\frac{\partial M_1}{\partial n_1}\bigg)_{T,\mu_0}=\bigg(\frac{\partial P_1}{\partial n_1}\bigg)_{T,\mu_0}.
\end{equation}
Batchelor derived his Eq. (6.9) for $\mu_{1B}$, not for $\mu_1$. However, his Eq. (6.9) reads like Eq. (4.8) with an equality sign.

Batchelor argued in his Eq. (6.1) that the thermodynamic force and the applied force should be equated as
\begin{equation}
\label{4.17}\frac{-\nabla\mu_1}{1-\phi_1}=\vc{E}_1.
\end{equation}
This leads to his expression for the diffusion coefficient which agrees with Eq. (3.3) with $\overline{v_1}$ replaced by $v_1$, as is correct for a Brownian suspension.
The agreement was noted for a Brownian system by Beenakker and Mazur \cite{21}. Batchelor's argument is reproduced by Russel et al. \cite{22}. Schurr \cite{6A} states that Batchelor based himself on Einstein's argument, but this is not the case.

To first order in $n_1$ by use of Eq. (4.13)
\begin{equation}
 \label{4.18}D_{GM}\approx D_0[1+(2B_2-6.55v_1)n_1]+O(n_1^2),
\end{equation}
with $D_0=k_BT/\zeta_0$ and sphere volume $v_1=(4\pi/3)a^3$, as in Batchelor's Eq. (6.12). Here we used Batchelor's virial correction to the Stokes friction coefficient $\zeta^*_{10}=\zeta_0 (1+6.55n_1v_1)+O(n_1^2)$, as evaluated from Stokes hydrodynamics with a no-slip boundary condition and a hard sphere pair distribution \cite{23}. In this case $2B_2=8v_1$.

In comparison Eq. (3.5) leads to
\begin{equation}
 \label{4.19}D_Y\approx D_0[1+(2B_2-\overline{v_1}-6.55v_1)n_1]+O(n_1^2).
\end{equation}
Berne and Pecora \cite{9} would have obtained this expression if they had calculated the thermodynamic derivative correctly, as noted below Eq. (4.13). Teraoka \cite{24} also arrived at an expression of the form Eq. (4.19), but he suggested a further term $-\overline{v_1}$ to account for backflow of the solvent. Such a term was also suggested by Chu \cite{25}. However, the term amounts to double counting, since solvent motion is accounted for in the calculation of the friction coefficient.

Instead, the generalized Einstein relation Eq. (2.14) leads to the virial expansion
\begin{equation}
 \label{4.20}D_{GE}=D_0[1+k_{DE}n_1]+O(n_1^2),
\end{equation}
with virial coefficient
\begin{equation}
 \label{4.21}k_{DE}=2B_2-k_fv_1,
\end{equation}
where $k_f$ is the virial correction to the friction coefficient $\zeta^*_{10}=\zeta_0[1+k_fv_1n_1]+O(n_1^2)$. For an incompressible viscous solvent the coefficient $k_f$ may be calculated from Stokes hydrodynamics for two spheres with a general boundary condition and pair distribution.

\section{\label{V}Generalized Smoluchowski equation}

Below Eq. (3.6) we questioned the force balance used in the derivation of Eq. (3.5). The force relation Eq. (4.17) used by Batchelor \cite{5} and by Russel et al. \cite{22} leads to the de Groot-Mazur expression Eq. (3.3). In this section we consider expressions derived for the collective diffusion coefficient $D_C$ of a suspension of interacting Brownian particles on the basis of the generalized Smoluchowski equation \cite{26}. The collective diffusion coefficient measures the decay of long wave fluctuations in the solute density.

The generalized Smoluchowski equation is based on a reduced description in which the solvent is treated as a continuum. The properties of the solvent are incorporated in effective direct interactions between solute particles and in a many-body mobility matrix derived from hydrodynamics. The Smoluchowski equation holds on a long time scale, long in comparison with the time of thermalization of the momenta of solute particles. The many-body Smoluchowski equation was first formulated in the theory of interacting Brownian particles by Deutch and Oppenheim \cite{10},\cite{11}.

In earlier work \cite{12} we found that the generalized Smoluchowski equation leads to a virial expansion of the form Eq. (4.20) with a virial coefficient $k_{DE}$ which agrees with the result derived by Batchelor \cite{5} for the case of hard spheres, though with an approximate value for $k_f=6.44v_1$ due to an approximate calculation of the pair hydrodynamic interaction. In work with Cichocki \cite{27} the coefficient was corrected to $k_f=6.546v_1$ in agreement with Batchelor \cite{23}. In later work we considered also the effect of more general effective pair interactions \cite{28}-\cite{31}. Similar results were derived by Van den Broeck et al. \cite{32},\cite{32A} on the basis of Einstein's argument for the thermodynamic virial correction and Batchelor's work on the frictional virial correction.

More generally, Smoluchowski's equation leads to an expression for the collective diffusion coefficient given by \cite{26},\cite{26A}
 \begin{equation}
 \label{5.1}D_C=D_0\lim_{k\rightarrow 0}\frac{H(k)}{S(k)},
\end{equation}
where $S(k)$ is the equilibrium structure factor of the solute with effective direct interactions as determined by the solvent at temperature $T$ and chemical potential $\mu_0$, and $H(k)$ is determined by hydrodynamic interactions and the equilibrium structure. The limiting value $S(0)$ is given by \cite{33}
 \begin{equation}
 \label{5.2}\frac{k_BT}{S(0)}=n_1\bigg(\frac{\partial M_1}{\partial n_1}\bigg)_{T,\mu_0}=\bigg(\frac{\partial P_1}{\partial n_1}\bigg)_{T,\mu_0}.
\end{equation}

Usually the generalized Smoluchowski equation is formulated for Brownian particles immersed in an incompressible viscous fluid, and correspondingly the diffusion matrix in the equation is related to the mobility matrix of Stokes hydrodynamics. However, the equation holds more generally. In particular, it can be applied to a collection of interacting particles immersed in a gas of much smaller molecules. The collisions with the gas of molecules are sufficient to cause Brownian motion of the big particles. The generalized Smoluchowski equation, with effective interactions calculated for constant $\mu_0$, leads to a collective diffusion coefficient given by the generalized Einstein relation Eq. (2.14).

The hydrodynamic factor $\lim_{k\rightarrow 0}H(k)$ was studied in both theory and computer simulation, incorporating Stokes hydrodynamics and an incompressible fluid. These calculations yielded interesting results in a wide range of densities for a variety of particle models, including hard spheres with no-slip boundary condition \cite{34} and porous spheres \cite{35}, in good agreement with theoretical calculations and computer simulations.

Phillies \cite{36},\cite{37} studied the first few terms of the virial expansion of $H(k)$, and claimed that an additional factor $1-\phi_1$ is needed to account for the transformation from the solvent-fixed frame to the laboratory-fixed frame, but in the light of the above this is not the case. In addition he claimed that the commonly accepted expression for the hydrodynamic factor $H(k)$ should be modified \cite{38},\cite{39}.

\section{\label{VI}Binary fluid mixture}

The two expressions for the mutual diffusion coefficient, $D_{GE}$ in Eq. (2.14) and $D_{GM}$ in Eq. (3.3), contain different thermodynamic derivatives. Hill \cite{40},\cite{41} studied the derivative at constant pressure from a statistical mechanical point of view in connection with the theory of light scattering and suggested that it may be usefully expressed in terms of a cluster expansion obtained in the pressure ensemble. It seems preferable to stick with the McMillan-Mayer theory of solutions \cite{42} and to use Vrij's thermodynamic identity Eq. (4.6) to relate the two derivatives. The derivative in the generalized Einstein relation Eq. (2.14) occurs naturally in the McMillan-Mayer theory.

As we have shown, for a nearly incompressible solution the two coefficients $D_{GE}$ and $D_{GM}$ are almost identical. The generalized Smoluchowski equation for a Brownian suspension holds also when the solvent is compressible. In this section we show that for a Brownian suspension the two expressions nearly agree, but that for a binary fluid mixture consisting of molecules of comparable size the two coefficients $D_{GE}$ and $D_{GM}$ can be quite different.

We cast the two expressions in a more transparent form by relating the thermodynamic derivatives to the Kirkwood-Buff correlation integrals \cite{43}, occurring in fluctuation theorems for density fluctuations. We write
 \begin{equation}
 \label{6.1}D_{GE}=\frac{S_1}{\zeta^*_{10}},\qquad D_{GM}=\frac{Q_1}{\zeta^*_{10}},
\end{equation}
with thermodynamic factors
 \begin{equation}
 \label{6.2}S_1=n_1\bigg(\frac{\partial\mu_1}{\partial n_1}\bigg)_{T,\mu_0},\qquad Q_1=\frac{n_1}{n_0\overline{v_0}}\bigg(\frac{\partial\mu_1}{\partial n_1}\bigg)_{T,p}.
\end{equation}
The factors can be expressed in terms of correlation integrals,
 \begin{equation}
 \label{6.3}G_{ij}=4\pi\int^\infty_0[g_{ij}(r)-1]r^2\;dr,
\end{equation}
 as
 \begin{eqnarray}
 \label{6.4}S_1&=&\frac{k_BT}{1+n_1G_{11}},\nonumber\\
 Q_1&=&k_BT\frac{n_0+n_1+n_0n_1(G_{00}+G_{11}-2G_{01})}{n_0[1+n_1(G_{11}-G_{01})]^2}.
\end{eqnarray}
In the second relation we have used the expressions given by Kirkwood and Buff for the thermodynamic derivative and for the partial volume $\overline{v_0}$.
From Eq. (6.4) the relation Eq. (4.5) can be read off immediately. Kirkwood and Buff also provide an expression for the compressibility $\kappa$ (with a misprint). By use of the Vrij identity Eq. (4.6) the osmotic compressibility $\kappa_{\pi 1}$ can be expressed as
 \begin{equation}
 \label{6.5}\kappa_{\pi 1}=\frac{1}{n_1Q_1}+\kappa=\frac{1}{n_1S_1},
\end{equation}
in agreement with Eq. (4.7).

We note that the ratio
 \begin{equation}
 \label{6.6}\frac{Q_1}{n_1\overline{v_1}^2}=\frac{k_BT}{n_0\overline{v_0}n_1\overline{v_1}}\big[n_0+n_1+n_0n_1(G_{00}+G_{11}-2G_{01})\big],
\end{equation}
with $G_{01}=G_{10}$, is symmetric in the labels $0,1$, so that
 \begin{equation}
 \label{6.7}\frac{Q_1}{n_1\overline{v_1}^2}=\frac{Q_0}{n_0\overline{v_0}^2},
\end{equation}
as we derived earlier from thermodynamics \cite{6}. The symmetry implies that the friction coefficient must possess the symmetry
 \begin{equation}
 \label{6.8}\zeta^*_{10}=n_1\overline{v_1}^2\psi,
\end{equation}
with a factor $\psi$ which is invariant under the interchange of labels if the diffusion coefficient $D_{GM}$ is to be invariant under the interchange. Similarly we find from Eq. (4.7)
 \begin{equation}
 \label{6.9}n_1\kappa_{\pi 1}S_1=n_0\kappa_{\pi 0}S_0=1.
\end{equation}

We present numerical results for a binary fluid mixture of hard spheres with radii $a_0$ and $a_1$ with thermodynamics calculated in Percus-Yevick approximation. The Helmholtz free energy of the mixture in volume $V$ was calculated by Lebowitz and Rowlinson \cite{44} (LR). The expression agrees with the virial expansions in Eqs. (4.1) and (4.2). We use the expression for the free energy as given in Eq. (6.1) of Ref. 6.

We consider isometric mixtures, defined by having equal volume fractions $\phi_0=n_0v_0$ and $\phi_1=n_1v_1$, where $v_j=(4\pi/3)a_j^3$. The total volume fraction is $\phi=\phi_0+\phi_1=2\phi_0$. The size ratio is characterized by the parameter
\begin{equation}
 \label{6.10}\Lambda=\log_{10}\frac{v_1}{v_0}=3\log_{10}\frac{a_1}{a_0}.
\end{equation}

In Fig. 1 we plot $\beta S_1$, where $\beta=1/(k_BT)$, as a function of volume fraction $\phi$ for three sizes $v_1=v_0/3,\;v_1=v_0$ and $v_1=3v_0$. We limit the volume fraction to values less than $0.6$. In Fig. 2 we plot $\beta Q_1$ as a function of volume fraction $\phi$ for the same size ratios. For large $\phi$ the functions $\beta S_1$ and $\beta Q_1$ tend to the same values, but there are significant differences at small volume fraction. We recall the limiting value found in Eq. (4.5). Also we note that for $v_1=v_0$ the correlation integrals satisfy $G_{00}=G_{11}=G_{01}=G_{10}$, so that then from Eq. (6.4) $Q_1=k_BT(1+x)$. In Fig. 3 we plot the ratio $D_{GE}/D_{GM}=S_1/Q_1$ as a function of volume fraction $\phi$ and size parameter $\Lambda$. The size parameter ranges from $-3$ to $3$.

We compared with results found from the approximate BMCSL expression for the free energy \cite{45}, which can be used alternatively \cite{46},\cite{47}. It yields results very similar to those in Figs. 1-3. The LR expression employed above requires less algebra and is somewhat easier to use. The third virial coefficients for mixtures of hard spheres are given by Erpenbeck \cite{48}, so that one can also compare with virial expansion results.

Fig. 3 makes clear that the two quantities $S_1$ and $Q_1$ are essentially identical in two important limit situations. The ratio $S_1/Q_1$ tends to unity at large volume fraction, where the mixture becomes incompressible, and at any volume fraction as the size parameter $\Lambda$ becomes large. The latter situation corresponds to the Brownian limit described by the Smoluchowski equation. The Brownian limit is sufficient for the identity of $S_1$ and $Q_1$, also when the suspension is compressible. We showed these properties for mixtures of hard spheres, but they clearly obtain more generally. It follows from Eq. (6.4) that the Brownian limit corresponds to the situation where $n_1<<n_0$ and the integrals $G_{00}$ and $G_{01}$ can be neglected in comparison with $G_{11}$.

\section{\label{VII}Kinetic theory}

The mutual diffusion coefficient can be studied from a more microscopic point of view in kinetic theory. Calculations based on the Boltzmann equation for a dilute gas mixture are reviewed by Chapman and Cowling \cite{49}. For dense hard sphere fluids van Beijeren and Ernst proposed a revised Enskog theory \cite {50}. The mutual diffusion coefficient of a binary mixture in an isobaric non-equilibrium situation takes the de Groot-Mazur form Eq. (6.1) with $Q_1$ expressed in terms of dimensionless quantities $E_{ij}$ defined as \cite{51}
 \begin{equation}
 \label{7.1}E_{ij}=\frac{n_i}{k_BT}\frac{\partial\mu_i}{\partial n_j}.
\end{equation}
In this notation
 \begin{equation}
 \label{7.2}Q_1=\frac{k_BT}{n_0\overline{v_0}}\;\frac{E_{11}E_{00}-E_{01}E_{10}}{E_{00}+E_{10}},
\end{equation}
with partial volume \cite{43}
\begin{equation}
 \label{7.3}\overline{v_0}=\frac{E_{00}+E_{10}}{n_0E_{00}+n_1E_{01}+n_0E_{10}+n_1E_{11}}.
\end{equation}
In the same notation from Eq. (2.9)
 \begin{equation}
 \label{7.4}S_1=k_BT\big[E_{11}-\frac{E_{01}E_{10}}{E_{00}}\big].
\end{equation}
The revised Enskog theory provides a calculation of the friction coefficient $\zeta^*_{10}$ in Eq. (6.1). Explicit calculations were performed by Kincaid et al. \cite{52}.

For particles with additional long-range interactions a mean-field kinetic theory was formulated by Karkheck et al. \cite{53},\cite{54}. The mutual diffusion coefficient of van der Waals binary mixtures was studied by Castillo et al. \cite{55}. The expression used is again of the de Groot-Mazur form with $Q_1$ given by Eq. (7.2). To compare with the result of the generalized Einstein relation it suffices to multiply by the factor $S_1/Q_1$.

\section{\label{VIII}Schurr's expression}

We return to Schurr's derivation mentioned in the Introduction. Schurr \cite{6A} used Einstein's argument to derive an expression for the diffusion coefficient which looks like the de Groot-Mazur expression Eq. (3.3). In his derivation he denoted the derivative $(\partial\mu_1/\partial p)_{n_1}$ as the partial volume $\overline{v_1}$. Instead we find in present notation
by use of the Gibss-Duhem relation
 \begin{equation}
 \label{8.1}\bigg(\frac{\partial\mu_1}{\partial p}\bigg)_{n_1}=\frac{1}{n_1}\;\frac{E_{10}}{E_{00}+E_{10}}=\overline{v_1}\;'.
\end{equation}
This differs from the expression for $\overline{v_1}$ found from Eq. (7.3) by an interchange of labels $(0,1)$, so that we introduce the new notation $\overline{v_1}\;'$.  A number of relevant thermodynamic relations is derived in appendix D to the article by Vafaei et al. \cite{42}.

From Eq. (7.2) and the expression Eq. (8.1) we find
 \begin{equation}
 \label{8.2}\frac{n_0\overline{v_0}}{1-n_1\overline{v_1}\;'}\;Q_1=S_1.
\end{equation}
This implies that the diffusion coefficients $D_{GE}$ and $D_{GM}$ are related as
 \begin{equation}
 \label{8.3}D_{GE}=\frac{n_0\overline{v_0}}{1-n_1\overline{v_1}\;'}\;D_{GM}=\frac{1}{\zeta^*_{10}(1-n_1\overline{v_1}\;')}\bigg(\frac{\partial\mu_1}{\partial n_1}\bigg)_{T,p}.
\end{equation}
The right hand side is the expression derived by Schurr \cite{6A}. However, note that $\overline{v_1}\;'$ is not a proper partial volume. The sum $n_0\overline{v_0}\;'+n_1\overline{v_1}\;'$ does not add up to unity, so that $n_1\overline{v_1}\;'$ cannot be identified with the volume fraction of solute, as done by Schurr. The relation Eq. (8.3) makes clear that the difference between the two diffusion coefficients corresponds to the difference between $\overline{v_1}\;'$ and $\overline{v_1}$.

\section{\label{IX}Generalized Stokes-Einstein relation}

The Stokes-Einstein expression $D_0=k_BT/(6\pi\eta a)$ for a dilute suspension of spheres of radius $a$ in a fluid of shear viscosity $\eta$ has led to the suggestion that at higher solute density the mutual diffusion coefficient can be expressed as the generalized Stokes-Einstein relation
 \begin{equation}
 \label{9.1}D_{SE}=\frac{k_BT}{6\pi\eta_{eff}a},
\end{equation}
where $\eta_{eff}$ is the effective viscosity of the suspension. Kholodenko and Douglas \cite{56} used mode-coupling theory to derive the modified expression
 \begin{equation}
 \label{9.2}D_{SEm}=\frac{k_BT}{6\pi\eta_{eff}\xi},
\end{equation}
where $\xi$ is a correlation length of the form
 \begin{equation}
 \label{9.3}\xi=aS(0)^{1/2}.
\end{equation}
The modified form Eq. (9.2) leads at low density to the wrong dependence on the thermodynamic virial coefficient. We regard the numerical agreement with $k_D$ for hard spheres as fortuitous.

Scott et al. \cite{57} suggested an expression for the diffusion coefficient corresponding to Eq. (4.18) with $D_0$ replaced by $D_{SE}$. This leads to the wrong virial expansion, and we presume that it also provides a poor approximation at higher density.

Banchio and N\"agele \cite{58} showed from Stokesian dynamics simulations that Eq. (9.1) provides a reasonably good approximation for a suspension of hard spheres over a wide range of solute density, even though the expression does not agree with the virial expansion Eq. (4.20). It may be worthwhile to compare Eq. (9.1) with Eqs. (2.14) and (3.3) for a variety of systems.

A recent application of Eq. (9.1) was made by Sorret et al. \cite{59}. In his monograph on diffusion Cussler \cite{60} mentions a number of semi-empirical relations, none of which is entirely successful.

\section{\label{X}Discussion}
In the above we considered three different expressions for the mutual diffusion coefficient of a binary fluid mixture, Eqs. (2.14), (3.3), and (3.5). The difference between the last two expressions is a reminder that irreversible thermodynamics is a tricky subject \cite{6A}.  The coefficient $k_D$ in the virial expansion of the diffusion coefficient which follows from Eq. (3.3) was verified experimentally for Brownian suspensions with nearly incompressible solvent \cite{9A},\cite{22}, and therefore the expression was accepted by many workers as the correct one.

For semi-dilute solutions the generalized Einstein relation Eq. (2.14) and the de Groot-Mazur expression Eq. (3.3) are identical to first order in solute density. The calculation of Sec. VI shows that for a binary fluid mixture of two species of molecules of comparable size the two predictions Eqs. (2.14) and (3.3) can be quite different, but it also makes evident that the two thermodynamic factors $S_1$ and $Q_1$ become essentially identical in two important limiting situations. When the mixture becomes incompressible the equality follows from Vrij's thermodynamic identity Eq. (4.6). The second limiting situation corresponds to a Brownian suspension. In that case the equality follows from Eq. (6.4) and the neglect of the solvent-solvent and solvent-solute correlation integrals with respect to the solute-solute integral.

It would be of interest to compare the two coefficients $D_{GE}$ and $D_{GM}$ for gaseous and liquid binary mixtures with molecules of comparable size in computer simulation and experiment. In recent molecular dynamics simulations \cite{62},\cite{63} the friction coefficient and the thermodynamic coefficient $Q_1$ were determined, and the de Groot-Mazur expression was used to calculate the diffusion coefficient $D_{GM}$. It would be desirable to determine the diffusion coefficient independently.

The generalized Einstein relation can be used with confidence for semi-dilute solutions, in the dense limit, and in the Smoluchowski limit. Einstein's argument is independent of the density and the particle size ratio. It is hard to see why it would break down in intermediate situations. The same can be said of the de Groot-Mazur derivation. It can be decided by experiment or computer simulation whether the generalized Einstein relation Eq. (2.14) or the de Groot-Mazur expression Eq. (3.3) is correct.\\\\

$\mathrm{\bf{Acknowledgment}}$ I thank Dr. J. Pathak for stimulating correspondence.\\\\

\newpage

\section*{Figure captions}

\subsection*{Fig. 1}
Plot of the function $\beta S_1$ as calculated in LR approximation for an isometric mixture of hard spheres as a function of total volume fraction $\phi=2\phi_0=2\phi_1$ for sizes $v_1=v_0$ (solid curve), $v_1=3v_0$ (long dashes), and $v_1=v_0/3$ (short dashes).

\subsection*{Fig. 2}
As in Fig. 1 for the coefficient $\beta Q_1$.

\subsection*{Fig. 3}
Plot of the ratio $S_1/Q_1$ as calculated in LR approximation for an isometric mixture of hard spheres as a function of total volume fraction $\phi=2\phi_0=2\phi_1$ and size parameter $\Lambda=\log_{10}(v_1/v_0)$.

\newpage
\setlength{\unitlength}{1cm}
\begin{figure}
 \includegraphics{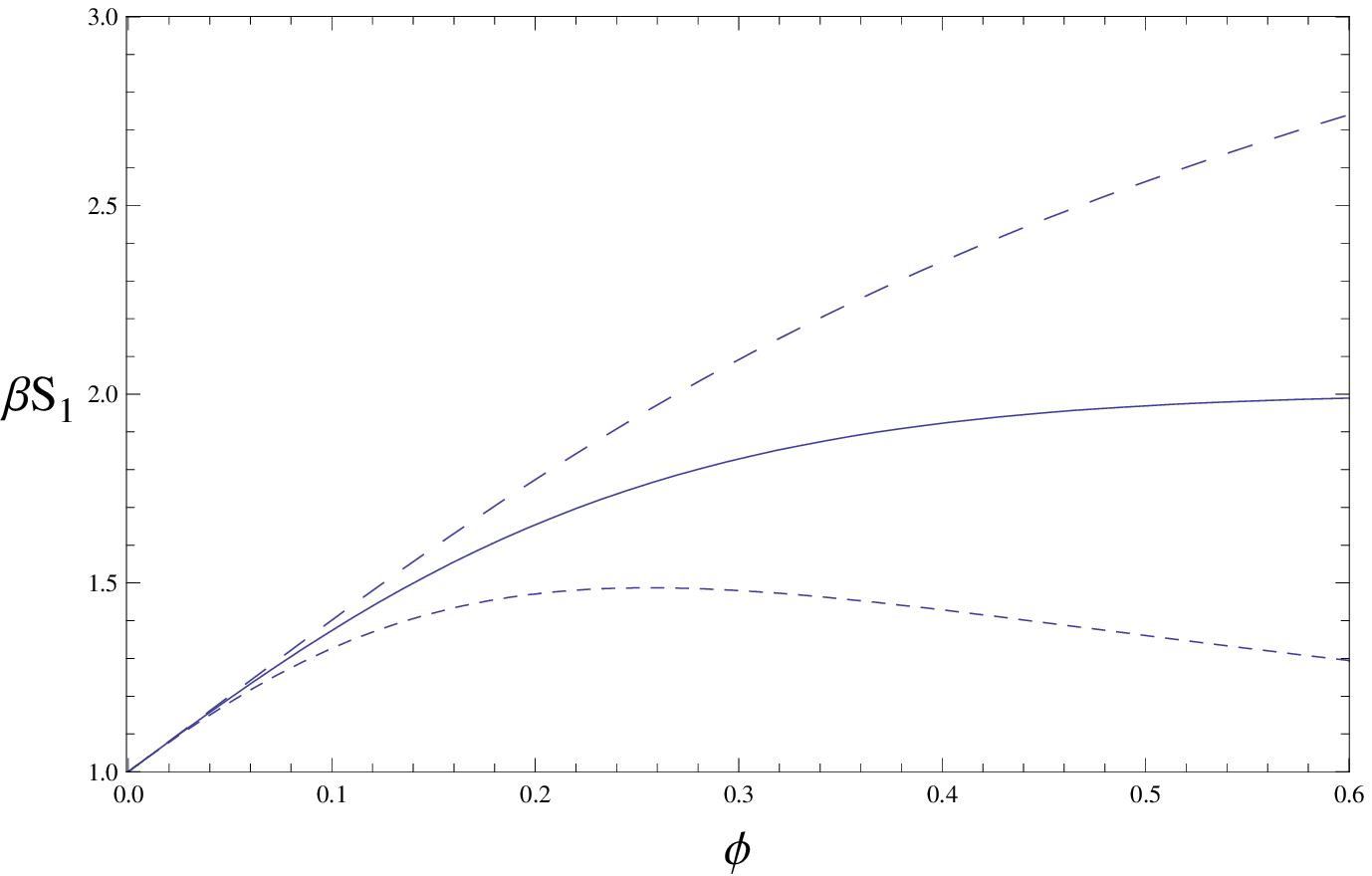}
   \put(-9.1,3.1){}
\put(-1.2,-.2){}
  \caption{}
\end{figure}
\newpage
\clearpage
\newpage
\setlength{\unitlength}{1cm}
\begin{figure}
 \includegraphics{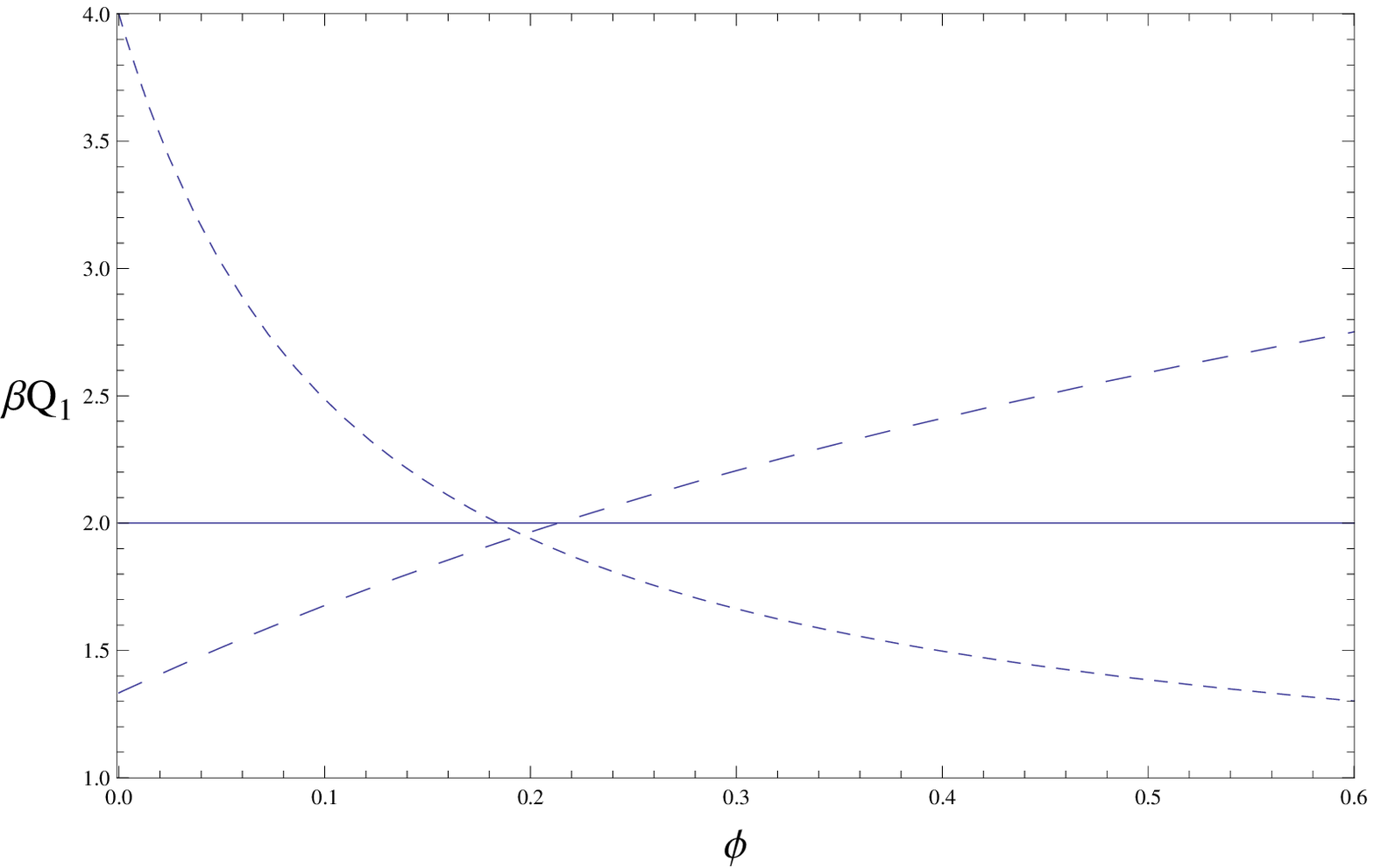}
   \put(-9.1,3.1){}
\put(-1.2,-.2){}
  \caption{}
\end{figure}
\newpage
\clearpage
\newpage
\setlength{\unitlength}{1cm}
\begin{figure}
 \includegraphics{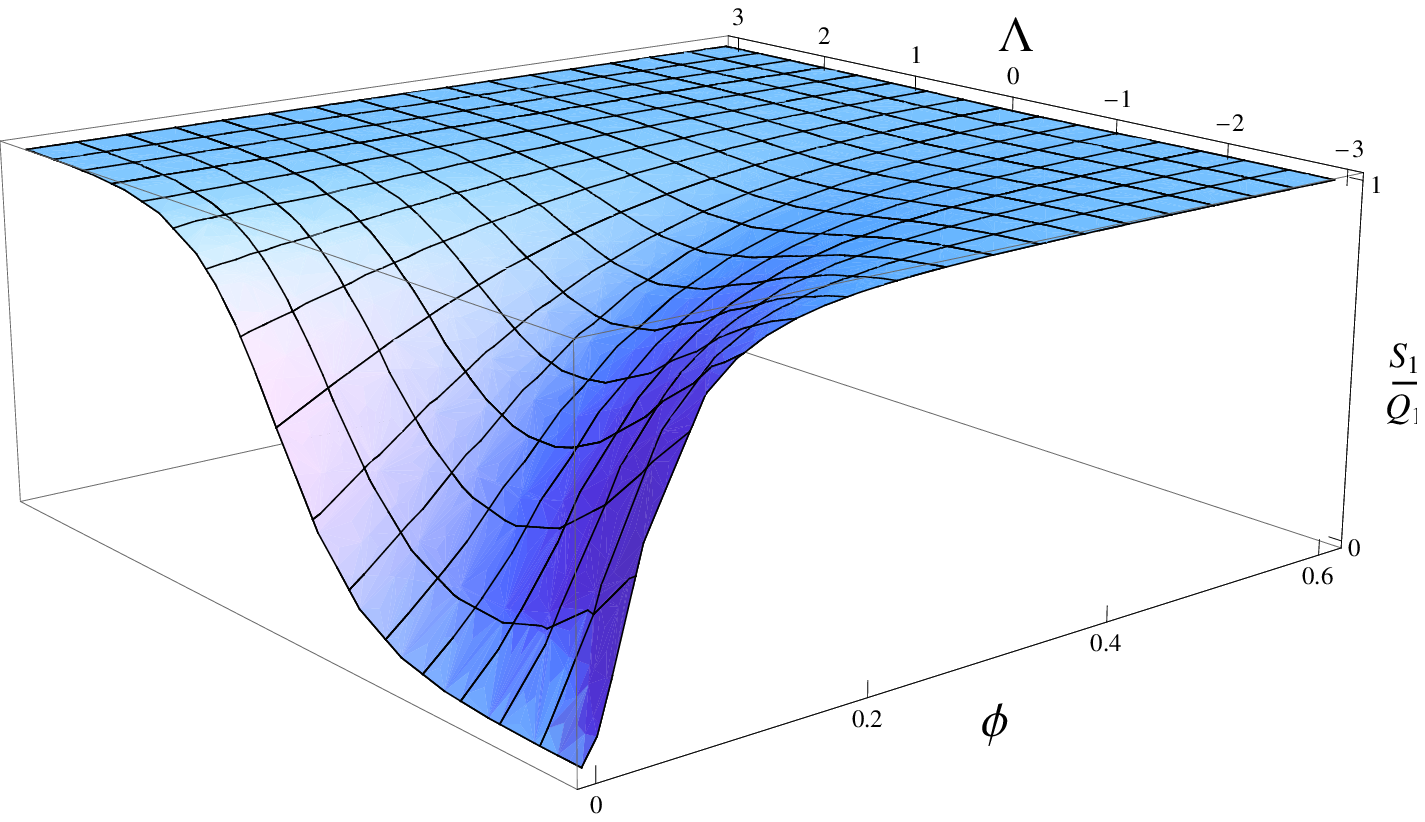}
   \put(-9.1,3.1){}
\put(-1.2,-.2){}
  \caption{}
\end{figure}
\newpage

\end{document}